\documentclass{article}

\usepackage[utf8]{inputenc} % allow utf-8 input
\usepackage[T1]{fontenc}    % use 8-bit T1 fonts
\usepackage{hyperref}       % hyperlinks
\usepackage{url}            % simple URL typesetting
\usepackage{booktabs}       % professional-quality tables
\usepackage{amsfonts}       % blackboard math symbols
\usepackage{nicefrac}       % compact symbols for 1/2, etc.
\usepackage{microtype}      % microtypography
\usepackage[final]{nips_2017} % produce camera-ready copy 

\title{Generating Black Metal and Math Rock:
Beyond Bach, Beethoven, and Beatles}

\author{
  Zack Zukowski\\
  Dadabots\\
  \texttt{thedadabot@gmail.com} \\
  \And
  Cj Carr \\
  Dadabots \\
  \texttt{emperorcj@gmail.com} \\
}

\begin{document}

\maketitle

\begin{abstract}
We use a modified SampleRNN [1] architecture to generate music in modern genres such as black metal and math rock. Unlike MIDI and symbolic models, SampleRNN generates raw audio in the time domain. This requirement becomes increasingly important in modern music styles where timbre and space are used compositionally. Long developmental compositions with rapid transitions between sections are possible by increasing the depth of the network beyond the number used for speech datasets. We are delighted by the unique characteristic artifacts of neural synthesis.

\end{abstract}

\section{Introduction}
The majority of deep learning papers on generative music focus on symbolic-domain generation, including all seven that appeared at ISMIR 2017[7][8][9][10][11][12][13]. Few have explored recent advances in neural synthesis (Wavenets[19], SampleRNN[1], DeepVoice[18]). Most style-specific generative music experiments have explored artists commonly found in harmony textbooks such as The Beatles[7], Bach[13], and Beethoven[1] but few have looked at generating modern genre outliers such as black metal.

The tradition of functional tonality in harmonic composition has been studied extensively for centuries and is taught in music theory courses today. But since the twentieth century, the study of manipulating timbre has played a much more significant role in composition technique. Composers like Varese thought in terms of composing "sound-masses" to construct his symphonic scores [6] inspiring future artists to discover new sonic material.

\section{Related Work}
NSynth is a promising approach to neural synthesis. Due to re-synthesis of the magnitude spectrum in the baseline model, phase retrieval artifacts are present, while models that predict samples in the time domain don't suffer from these artifacts[2].

\section{Our Process}

We pre-process each audio dataset into 3,200 eight second chunks of raw audio data (FLAC). The chunks are randomly shuffled and split into training, testing, and validation sets. The split is 88\% training, 6\% testing, 6\% validation.

We use a 2-tier SampleRNN with 256 embedding size, 1024 dimensions, 5 to 9 layers, LSTM or GRU, 256 linear quantization levels, 16kHz sample rate, skip connections, and a 128 batch size, using weight normalization. The LSTM gated units have a forget gate bias initialized with a large positive value of 3. The initial state h0 is either learned or randomized. We train each model for about three days on a NVIDIA K80 GPU. 

Intermittently at checkpoints during training, audio clips are generated one sample at a time and converted to a WAV file. Originally SampleRNN used an argmax inference method. We modified it to sample from the softmax distribution.

At each checkpoint we generate 10x 30 second clips.  Early checkpoints produce generalized textures. Later checkpoints produce riffs with sectional transitions. If after a few epochs it only produces white noise, restart the training.

Sometimes a checkpoint generates clips which always get trapped in the same riff. Listen for traps before choosing a checkpoint for longer generations. 

The number of simultaneously generated clips (n\_seq) doesn't effect the processing time, because they are generated in parallel. The number is limited by GPU memory. 

\section{Results}

Our best results used 5-layer LSTM, trained on whole albums, for 50k-80k iterations. Randomizing the initial state h0 generates more varied music. Datasets with a unified sound (songs with similar instrumentation that were mixed and mastered together) were better able to generalize and combine ideas together. 

Audio examples are available here: dadabots.com/nips2017/

\subsection*{Room For A Ghost: mixed meter, odd time signatures, and abruptly sectional transitions}

We generated six minute compositions from a three song album by the rock band Room For A Ghost using 7-layer LSTM. The result retained the timbre of the original band, but had become "math rock". There were abrupt sectional changes, odd meters, and long rests. Math rock favors disjointed melodic contours with distorted tone and shifting metric emphasis.  The original music did not have these elements. 

\subsection*{Krallice: atmospheric texture, tremolo-picked guitars, slow weaving sections}

Krallice is an American black metal project. The style is characterized by its ultra long  progressive sections, textural rhythms, deep screams, and melodic weaving over a grid of steady, aggressive rhythmic attacks. These extreme characteristics make it an outlier in human music. 

We preprocessed 35 minutes audio taken from a single album. We trained using 5-layer LSTM and GRU models. The GRUs failed to learn the audio resulting in harse noise when sampled. The LSTMs were successful in training and sounded like Krallice. We generated twenty sequences with four minute durations.

\subsection*{Aesthetics of Neural Synthesis}

While we set out to achieve a realistic recreation of the original data, we were delighted by the aesthetic merit of its imperfections. Solo vocalists become a lush choir of ghostly voices, rock bands become crunchy cubist-jazz, and cross-breeds of multiple recordings become a surrealist chimera of sound. Pioneering artists can exploit this, just as they exploit vintage sound production (tube warmth, tape-hiss, vinyl distortion, etc).

\section{Conclusion and Future Work}

Creatively, we emphasize the importance of building neural synthesis models capable of generating music from raw audio, beyond just symbolic representations. Future work includes exploring local conditioning with hybrid representations of raw and symbolic audio.

\section*{References}

\small

[1]  Soroush Mehri, Kundan Kumar, Ishaan Gulrajani, Rithesh Kumar, Shubham Jain, Jose Sotelo, Aaron C. Courville, and Yoshua Bengio, (2017) SampleRNN: An Unconditional End-to-End Neural Audio Generation Model

[2] Jesse Engel, Cinjon Resnick, Adam Roberts, Sander Dieleman, Douglas Eck, Karen Simonyan, and Mohammad Norouzi, (2017) Neural Audio Synthesis of Musical Notes with WaveNet Autoencoders

[3] Ian Simon, and Sageev Oore, (2017) blog: https://magenta.tensorflow.org/performance-rnn

[4] Louis Bigo, Mathieu Giraud, Richard Groult, Nicolas Guiomard-Kagan, and Florence Leve, (2017) MIDINET: A Convolutional Generative Adversarial Network For Symbolic-Domain Music Generation

[5] Junyoung Chung Caglar Gulcehre KyungHyun Cho Yoshua Bengio
Empirical Evaluation of Gated Recurrent Neural Networks on Sequence Modeling: arXiv:1412.3555v1 [cs.NE] 11 Dec 2014

[6] Chou Wen-chung. 1966b. 
"Varèse: A Sketch of the Man and His Music". The Musical Quarterly 52, no. 2 (April): 151–170.

[7] François Pachet, Pierre Roy, Alexandre Papadopoulos (2017)
Sampling Variations of Sequences for Structured Music Generation

[8] Li-Chia Yang, Szu-Yu Chou, Yi-Hsuan Yang,  (2017) MidiNet: A Convolutional Generative Adversarial Network for Symbolic-Domain Music Generation

[9] Kosetsu Tsukuda, Keisuke Ishida, Masataka Goto,  (2017) Lyric Jumper: A Lyrics-Based Music Exploratory Web Service by Modeling Lyrics Generative Process

[10] Shunya Ariga, Satoru Fukayama, Masataka Goto,  (2017) Song2Guitar: A Difficulty-Aware Arrangement System for Generating Guitar Solo Covers from Polyphonic Audio of Popular Music

[11] Hyungui Lim, Seungyeon Rhyu, Kyogu Lee,  (2017) Chord Generation from Symbolic Melody Using BLSTM Networks

[12] Yifei Teng, Anny Zhao, Camille Goudeseune,  (2017) Generating Nontrivial Melodies for Music as a Service

[13] Hadjeres, G., Pachet, F. and Nielsen, F.,  (2017) DeepBach: a Steerable Model for Bach Chorales Generation.

[14] Marchini, M., Pachet, F. and Carré, B.,  (2017) Rethinking Reflexive Looper for structured pop music.

[15] Eigenfeldt, Arne \& Bown, Oliver \& Brown, Andrew \& Gifford, Toby. (2016). Flexible Generation of Musical Form: Beyond Mere Generation. 

[16] Hang Chu, Raquel Urtasun, Sanja Fidler
Song From PI: A Musically Plausible Network for Pop Music Generation

[17] http://www.flow-machines.com/ai-makes-pop-music/

[18] Sercan O. Arik, Mike Chrzanowski, Adam Coates, Gregory Diamos, Andrew Gibiansky, Yongguo Kang, Xian Li, John Miller, Andrew Ng, Jonathan Raiman, Shubho Sengupta, Mohammad Shoeybi, (2017) Deep Voice: Real-time Neural Text-to-Speech

[19] Aaron van den Oord, Sander Dieleman, Heiga Zen, Karen Simonyan, Oriol Vinyals, Alex Graves, Nal Kalchbrenner, Andrew Senior, Koray Kavukcuoglu, (2016) WaveNet: A Generative Model for Raw Audio

[20] John M. Chowning Papers (SC0906). Department of Special Collections and University Archives, Stanford University
Libraries, Stanford, Calif.
\end{document}